# The 1908 Tunguska event and the atmospheric optical anomalies

Andrei Ol'khovatov

https://orcid.org/0000-0002-6043-9205

Independent researcher
(Retired physicist)

Russia, Moscow
email: olkhov@mail.ru

submitted to arxiv.org

**Dedicated to the blessed memory of my grandmother ( Tuzlukova Anna Ivanovna ) and my mother ( Ol'khovatova Olga Leonidovna )**

**Abstract.** This paper is a continuation of a series of works, devoted to various aspects of the 1908 Tunguska event. In the late June - early July, 1908 there were various optical anomalies in the atmosphere. Early the Author already considered "bright nights", and in this paper some other atmospheric optical anomalies are considered, such as prolonged twilights, etc. Much attention is paid to the review of works on the study of atmospheric transmission at the Smithsonian Astrophysical Observatory in 1908. The main features of the atmospheric optical anomalies and their consequences for the interpretations of the Tunguska event are described. It is shown that neither the asteroidal nor the cometary interpretation of the Tunguska event is consistent with the anomalies. The geophysical interpretation can explain the atmospheric optical anomalies at the qualitative level at least. Quantitative explanation is currently difficult due to limited observational data and the lack of clarity in the details of physical phenomena.

## 1. Introduction

This paper is a continuation of a series of works in English, devoted to various aspects of the 1908 Tunguska event [Ol'khovatov, 2003; 2020a; 2020b; 2021; 2022; 2023a; 2023b; 2025a; 2025b; 2025c; 2025d; 2025e; 2025f]. The works can help

researchers to verify the consistency of the various Tunguska interpretations with actual data. A large number of hypotheses about its causes have already been put forward. However, so far none of them has received convincing evidence.

In the late June - early July, 1908 there were various optical anomalies in the atmosphere. In [Ol'khovatov, 2025a] the phenomenon called "bright nights" was considered. In this paper some other anomalies are considered.

Let's start with brief info about research of the Tunguska event. The Committee on Meteorites of the USSR Academy of Sciences (KMET) stopped research the area of the Tunguska event in the early 1960s. Later amateurs most of whom united under the name Kompleksnaya Samodeyatel'naya Ekspeditsiya (KSE) continued research. Since the late 1980s foreign scientists take part too.

Please pay attention that so called the epicenter of the Tunguska forestfall (the forestfall is named "Kulikovskii") is assigned to 60°53' N, 101°54' E, and in this paper is called the epicenter.

In this paper its author (the author of this paper i.e. A.O.) for brevity will be named as "the Author". The surname Vasil'ev can also be translated as Vasilyev in some references.

## 2. The atmospheric optical anomalies

In general there were various optical sky peculiarities at those times. Krinov wrote [Krinov, 1966]:

"On the first night after the fall of the Tunguska meteorite, i.e. from 30 June to 1 July 1908, and with lesser intensity on a few successive nights, extraordinary optical phenomena were observed in the Earth's atmosphere."

However the phenomena commenced before June 30. Here is from [Vasilyev, 1998]:

"A detailed analysis of dynamics of June-July 1908 optical anomalies gives ground to a supposition that their first signs had appeared already several days before the fall of the meteorite: Suring (1908) points out that they began on 23 June, F. de Roy (1908) about 25 June and Denning (1908a, b) on 29 June. On that last day they were registered at 8 points in Germany, Holland, Britain, Sweden, Poland and Russia (Fig. 6). Nevertheless, the events reached their maximum (observed at more than 140 points) on the night of 30 June to 1 July.<…> Beginning with 1 July they vanished exponentially, but post-effects continued until the end of July 1908."

In [Ol'khovatov, 2025a] some examples of the anomalies before June 30, 1908 are presented. The anomalies reached their peak on the evening of June 30th.

The late evening of June 30, 1908 was remembered by eyewitnesses, first of all, because of the colorful twilight, which in some places lasted all night. So D.D. Rudnev in the village of Muratovo (the Oryol province) captured the twilight with his photocamera at 0:30 on July 1, and then manually colored the picture. The shutter speed was 10 minutes. This picture is shown in Fig.1 taken from [Nadein, 1909]. Since Muratovo is a very common name for a settlement, the Author can only say that it is most likely the village of Muratovo with coordinates about 53.2° N, 35.8° E.

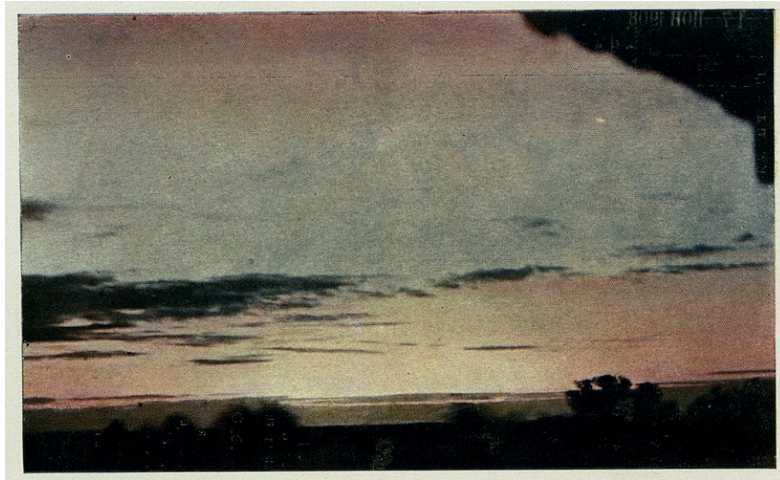

**Fig.1**

In his publication V.G. Fesenkov wrote [Fesenkov, 1949] (translated by A.O.):

"The fall of the famous Tunguska meteorite caused extraordinary atmospheric phenomena, which manifested themselves in a sharp increase in the glow of the night sky. There were bright dawns and extremely long twilight periods, which were observed over vast areas of the globe. These phenomena (by their nature) must be caused by the large amounts of finely dispersed matter that were released into the atmosphere. Therefore, it seems highly likely that the transparency of the Earth's atmosphere was also reduced particularly for short wavelengths. The only station on the Globe on which regular determinations of the transparency coefficient for different waves began at that time in connection with the investigation of the value of the solar constant, was at a very great distance from the place of the fall, on Mount Wilson in California. <...>

By examining these data |1| and comparing them with similar readings at the same station and by the same methods in the next few years, it is possible to state a noticeable decrease in the transparency of the

atmosphere, beginning approximately in the middle of July 1908 and up to the middle of August. <...>

By plotting these values on a graph (Figure), we can see that there is a significant decrease in atmospheric transparency from the second half of July 1908, which is significantly different from the three subsequent years. <...>

It is highly likely that this phenomenon is caused by the dispersion of the Tunguska meteorite."

Fesenkov plotted the average values of the transparency coefficients for the pentads of June-September from 1908 to 1911 for three different wavelengths (0.7 μm, 0.45 μm, 0.4 μm), which are shown in Fig.2 (adapted from [Fesenkov, 1949]).

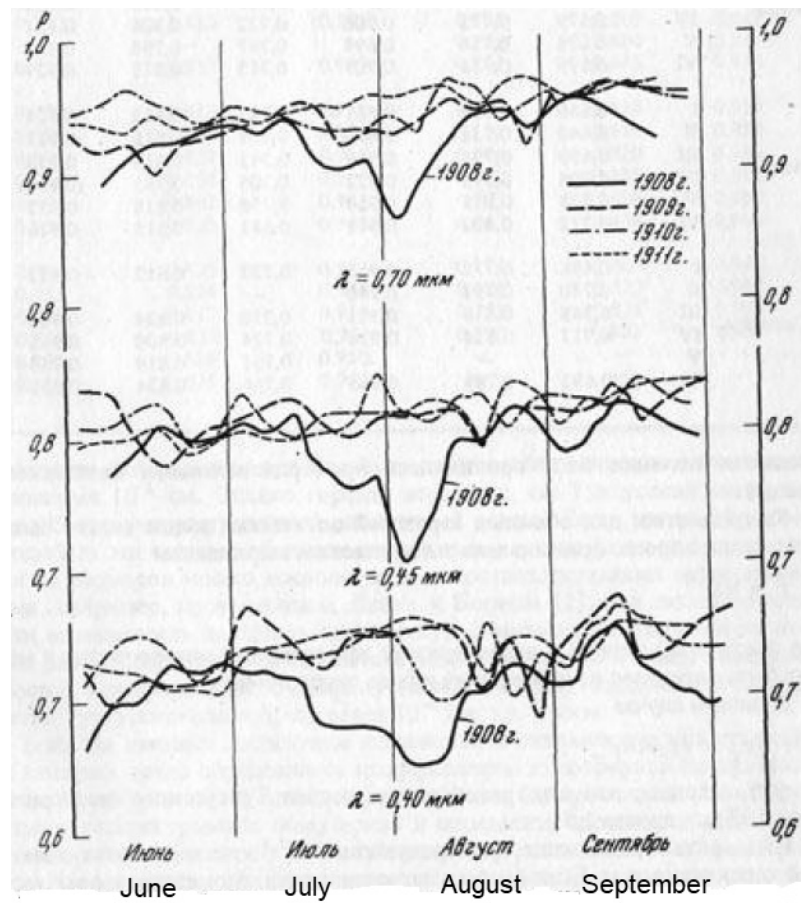

**Fig.2**

Fesenkov ended his article [Fesenkov, 1949] with words (translated by A.O.):

"However, the fact that the turbidity/opacification that reached California two weeks after the catastrophic fall of the meteorite lasted for about a month, and that it thus spread over a vast area of the Earth's surface, allows us to

estimate the order of magnitude of the mass of this meteorite to be at least several million tons."

Unfortunately, Fig.2 does not show the transparency values for the beginning of June 1908. This data is presented in much more detailed article [Turco et al., 1982]. Here is a couple fragments from its abstract:

"Atmospheric transmission data collected by a research team of the Smithsonian Astrophysical Observatory (APO) at Mount Wilson, California, from 1908 to 1911 are analyzed for ozone absorption in the Chappuis bands. Statistical analysis of the APO data reveals an ozone variation of 30 ± 15% over this period, supporting the theoretical predictions. <...> The chemical afterglows are shown to be intense enough to account for some of the unusual night-time light displays seen after the fall, but not widespread enough to explain the "light nights" and glowing skies reported throughout Eurasia. These phenomena appear to be related to the dust and water vapor deposited by the meteor at the cold summer mesopause, resulting in the formation of dense noctilucent clouds. Only circumstantial optical evidence for a large Tunguska $NO_2$ enhancement is found, which can not be used to calibrate independently the NO injection by the meteor. The suggestion of a dust veil created by the Tunguska explosion is revealed by the APO transmission data. We deduce that nearly 1 million tons of pulverized dust may have been deposited in the mesosphere and stratosphere by the Tunguska fall, which agrees with previous estimates of the meteor mass influx. Possible climate changes triggered by the Tunguska event are investigated. The most important climate anomaly identified in the post-Tunguska era is a 0.3°K cooling of the Northern Hemisphere which lasted for almost a decade. Several large volcanic eruptions occurred during this period which also played a role in the temperature change. However, radiation transport calculations are reported which suggest that Tunguska contributed to the cooling trend."

The authors of [Turco et al., 1982] considered that the Tunguska spacebody weighed about $5 \times 10^6$ tons, and most of this material consisted of $H_2O$, $NH_3$, $CH_4$, and $CO_2$ ices. So they wrote that an estimate of $\sim 1 \times 10^6$ tons of meteoric dust deposited in the upper atmosphere appears reasonable.

The authors of [Turco et al., 1982] made more detailed plot with more wavelengths and wrote that it appears that, for the entire year of 1908, the atmosphere was in a highly disturbed state. They marked another significant increase in atmospheric opacity which occurred 2 weeks prior to the Tunguska event (and was not shown of the Fesenkov's plot), but they added that this may have been associated with a dust cloud remnant from the Russian volcano of 1907. They explained the

latter in the following way [Turco et al., 1982]:

"The June and July/August anomalies, when compared in Fig. 7, show quite different wavelength dependences. The June anomaly exhibits a strongly decreasing optical depth in the near-ultraviolet region. This behavior is difficult to explain if a highly (size) dispersed dust cloud is responsible. Thus, while the June anomaly might have been caused by a remnant of the Shtyubelya Sopka eruption cloud formed 15 months earlier, it more likely had another origin. By contrast, the July/August turbidity spectrum closely corresponds to that seen shortly after the Katmai eruption (Volz, 1975). This suggests a relatively fresh dust cloud that might be associated with the Tunguska meteor."

In 1988 a new article was published which proposed another interpretation (but still in the frame of the Tunguska spacebody infall) [Kondratyev et al., 1988]. Its leading author was academician Kirill Yakovlevich Kondratyev ( https://new.ras.ru/staff/akademiki/kondratev-kirill-yakovlevich/ ). Here are some of the main points from [Kondratyev et al., 1988] in a simplified form. The Author did his best to avoid possible distortions of the meaning, but please consider it with some caution (TS is the Tunguska spacebody):

1. The turbidity of the atmosphere caused by the TS is not due to dust from the body's entry and explosion. The TS did not introduce much dust into the atmosphere, which explains the short duration of optical anomalies in Eurasia after its entry.

2. The increase in the optical density of the atmosphere in September and October is not due to the passage of TS products over Mount Wilson, but, like the cloud in May and June, is caused by a dust cloud formed in the stratosphere by another space object.

3. The spectral dependence of the averaged value of the aerosol component of the optical density in May-October 1908 does not approximate the dependence obtained in [Fesenkov, 1949] and does not correspond to the dependence presented in [Turco et al., 1982] throughout the entire studied spectral range. It is not possible to use the blue-red ratio on 0.4 microns wavelength to calculate the spectral dependence of the aerosol component of the optical density in the situation.

4. As a background (i.e., unperturbed) value of the optical density, instead of the average value for the period of years 1909-1911, it is better to use the average value for the year 1911, as the atmosphere had become more completely cleared (restored to its optical properties) after the disturbances of the year 1908.

5. It is not justified to refuse to calculate the optical density components for 1908 directly from the measurements of that year and to use the method of indirect determination by extrapolating the data for the period of 1909-1911.

6. Due to the fact that the dependence of the optical density of the aerosol component on the wavelength has a complex shape in the spectral region (0.4-0.8 μm), a linear approximation cannot be used to determine the residual optical density of the ozone component and the concentration of ozone, as was done in [Turco et al., 1982 ].

The authors [Kondratyev et al., 1988] also presented a graph (based on the data from the Smithsonian Astrophysical Observatory) of the time course of the total optical density of the atmosphere (for wavelengths of 700 nm and 1000 nm) and the residual optical density (for a wavelength of 400 nm). The graph (adapted by A.Yastrebov in color) is shown on Fig.3.

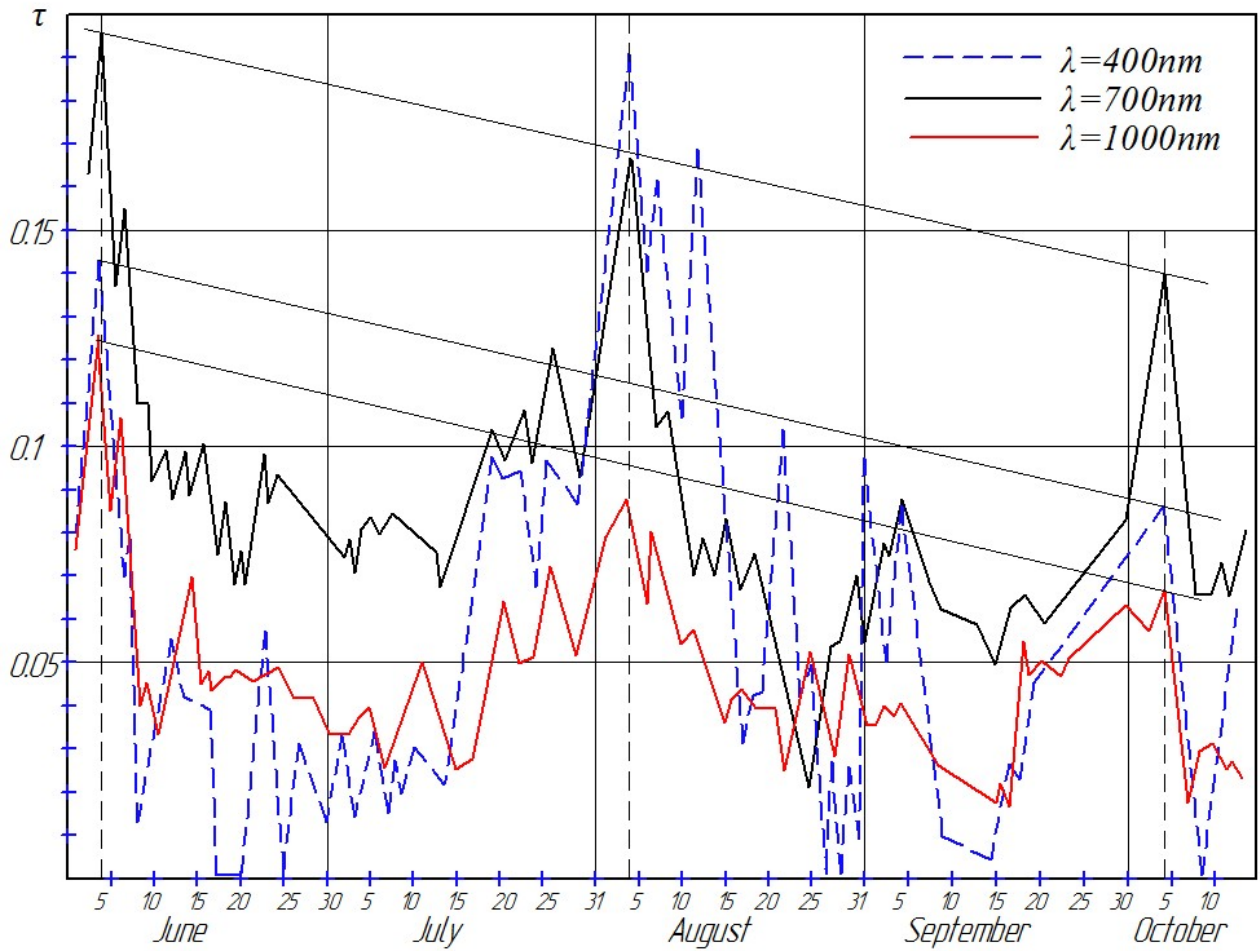

**Fig.3**

According to the authors [Kondratyev et al., 1988], at a wavelength of 400 nm, there was an overlap of the optical density associated with the TS cloud with the optical density of the dusty air mass at an altitude of 21-27 km, which circulated around the globe with a period of 60 days.

As an intermediate result, the authors [Kondratyev et al., 1988] write the following, translated by A.O.:

"Thus, the dust cloud that first passed through Mount Wilson on June 4,

1908, was formed and spread in the area where high-speed bolides stopped, and from our perspective, it represents the debris of a large bolide. By extrapolating the erosion of the cloud in time and the velocity of its spread, it can be assumed that the bolide explosion occurred over the Pacific Ocean, north-east of the Kuril Islands. The mass of the bolide is estimated by order of magnitude at 0.1 Mt."

The authors of [Kondratyev et al., 1988] estimated the initial mass of TS as ~70-250 Mt. There are several other ideas in the article, but they are not considered here. The Author wants to encourage interested readers to read the article [Kondratyev et al., 1988] to get the most accurate understanding of the topic.

The next article was published without K.Ya. Kondratyev as an author [Nikol'skii and Shul'ts, 1990]. The authors write that they considered extended time interval 1905-1913 and have improved some methods. According to the authors, TS did not contribute much dust to the atmosphere and consisted almost entirely of water-ammonia-methane ice and other components predicted by modern theories of comet nuclei. The authors estimated the mass of water in TS (the average excess total moisture content 0.8 $cm^{3\ [sic!]}$) at 1700 Mt [sic!], based on the increase in moisture content. So the total mass is 1.5-2 times greater [Nikol'skii and Shul'ts, 1990]. The authors noted that starting from June 19, there was a linear increase in the minimum values of the blue-red ratio over time, which may indicate a proportional accumulation of water vapor in the upper atmosphere. The authors point out that this phenomenon is not observed in the data from other seasons they have studied. However, in the opinion of the authors, this observation is consistent with the anomalous optical phenomena observed in June 1908. The general increase in the moisture content of the atmosphere in its linear growth led to a value of 2.6 cm of precipitated water at the end of this process (around August 8) [Nikol'skii and Shul'ts, 1990]. This overmoistening of the atmosphere, especially its upper layers, should have caused atmospheric optical anomalies, according [Nikol'skii and Shul'ts, 1990].

In the article published in 2012 [Nikol'skii et al., 2012] the authors wrote regarding the article [Nikol'skii and Shul'ts, 1990], translated by A.O.:

"These well-founded hypotheses turned out to be sufficient to explain in a consistent and complete way the whole complex of events that preceded the catastrophe, constituted its essence, and followed it. Their complex is systemically and consistently considered in the publication [Nikol'skii and Shul'ts, 1990]. It should be noted right away that due to the turmoil in the editorial board of "Meteoritika", which occurred during the editing of this publication, we did not receive the proofs of the article, and the new editor, by arbitrarily manipulating the text, drawings, tables, dimensions, etc., made a number of inconsistencies and, to a certain extent, devalued this work, making it difficult to read. Over the past time our views and estimates in some points have undergone changes, which were reported by us at several conferences with conjugate themes and were published in their works [Nikol'skii et al., 1997]."

Fortunately there is a conference report from 2009 in English [Nikolsky et al., 2010] (please pay attention to a bit different translated surname's spellings). In the report the researchers tried to develop the cometary interpretation to explain their results. Here are several fragments from [Nikolsky et al., 2010] with an accent on the atmospheric optics (TCB is the Tunguska Comet Body, and TE is the Tunguska Event):

"As a base for the construction of a picture of Tunguska event, we accept the following reasonable hypotheses: 1) bodies intruded into the Earth atmosphere formed as a result of past disintegration of a comet nucleus (possibly — Encke–Baklund); 2) the mass of the intruded fragments was significant ≥ 32 Mt, 3) the body was captured into a satellite orbit; 4) in the final stages the body was destroyed and its fragments had rather low speed (~ 2 km/s); 5) the basic source of the devastating energy was a volume explosion of a detonating gas-air mixture of hydrocarbon (CH) components of cometary substances with oxygen of the air. <...>

The total mass of the comet substance interacting with the atmosphere during the Tunguska catastrophe was estimated by us from the data of the Smithsonian Astrophysical Observatory (SAO) on atmosphere spectral transparency [2]. At the moment of catastrophe the observatory on Mount Wilson was the only one in the world where spectral measurements of atmospheric transparency were taken. These and some other data make it possible to state that, firstly, the post explosion air mass had reached Mount Wilson and, secondly, some of the registered deviations in the SAO data were caused by the catastrophe. Corrections to repeatable errors have been estimated for the SAO data; a novel method has been developed to separate a spectral optical thickness into components in absorption bands; to estimate the TCB mass (V. G. Fesenkov, R. Ganapahty) and the ozone losses (R. Turko) [2, 3], the errors and their origins have been determined. It was found that somewhere between the end of April and beginning of May a tandem of three large bodies behaving like carbon-bearing chondrites intruded into the Earth's atmosphere and exploded at the height of the ozone layer maximum. They formed an optically dense dust cloud in the stratosphere at altitude of 22–27 km, which went over the observatory three times with a period of 60 days. It is from this cloud that V. G. Fesenkov estimated the mass of dust as 1 Mt. We estimated the value as about 0.1 Mt. When going over the observatory at the second time, the cloud optically superposed the post explosion air mass; the mass was moving at an altitude of about 16–19 km, hardly had any dust, and was rich in water and nitrogen oxides. Fig. 2 represents the variations of spectral transparency, total water content and total ozone content which concordantly support the dynamic and optical processes registered in the Northern hemispheric stratosphere while driving the post explosion air mass from Eurasia to North America. The water content remained excessive (0.8 cm in the average) from July 15th till August 12th, which supports not only the time (fourteen 24-hour periods) taken by the post explosion air mass to reach Mount Wilson, but also the fact that the air stream in the mid-latitude Ferrell cell drew the products of ablation and of volumetric explosions from the catastrophe region during twenty-eight 24-hour periods. Note that in June of the last decade before the TE, experts observed a solar halo in a quite cloudless sky indicating that the Earth entered the zone of daytime β-Taurids meteor shower, the stream reaching maximum on June 30th, 1908. The southern trajectory of the TCB intrusion, which the authors have accepted, is sustained by log-book notes of the British Antarctic expedition that was stationed in vicinity of the magnetic pole in summer 1908. The notes will lead one to another conclusion: the TCB intrusion had a low impact parameter; the body significantly slowed down, intensively broke up and formed plasma; the plasma extended up into the ionosphere and may have had a global impact on the geomagnetic field."

The last more or less detailed article of the authors regarding the Tunguska event appeared in 2012 [Nikol'skii et al., 2012]. It should be noted that the authors tended to write very long and complicated sentences, the meaning of which the Author could not always grasp easily. Anyway, here are several fragments translated by A.O. (TM is the Tunguska meteoroid, TKT is the Tunguska cometary body):

"However, a more thorough and in-depth analysis of the spectral data we found in the Annals of the Smithsonian Observatory allowed us to assert that two months before the fall of TM, there was an invasion of a high-speed group of 3 bodies, which were apparently chondrite meteorite bodies, which we called the "Pre-Tunguska meteoroid" (PTM). They entered the Earth's atmosphere in late April or early May. <...> The "Pre-Tunguska meteoroid" was, judging by the optical thickness and the space-time structure of the dust layer it formed, a cluster of three large bodies of the type of carbonaceous chondrites (with a total mass of ~ 0.3 Mt, according to the latest estimates), which were completely or partially destroyed in the stopping zone (in the layer 22-27 km) [Nikolskii, 1985]. The dust clouds formed by them have a period of approximately 60 days and have at least twice circled the globe [Kondratyev et al., 1988; Nikol'skii and Shul'ts, 1990]. The optical thickness of these dust clouds, which was formed (in the optical sense) during their second passage over the Mount Wilson Observatory with the post-catastrophic TKT air mass, was estimated (based on the blue-red ratio) by V. G. Fesenkov at 1 M [Fesenkov, 1949]. <...>.

The next fundamental factor of the proposed model is the capture hypothesis of TKT by the Earth's gravitational field into the orbit of a satellite with a short aiming distance (low perigee) and, accordingly, with intense deceleration during each of its 4 passes through the perigee at altitudes from 40 to 36 km.  <...>

At an altitude of 24 km, the body has the following dynamic parameters: a velocity modulus of 6.5 km/s, an azimuth of 181° and an angle of inclination to the surface of 7°; after passing 4 perigees, the radius of the body decreases up to 92 m (m = 3.26 Mt). Detailed modeling of the separation of this body and its subsequent immersion into the atmosphere requires clarifying the initial data on the composition and structure of the TM. It is obvious that such information will never become our property, since the TKT has undergone processes of ablation, fragmentation, evaporation, and subsequent fragmentation into blocks (partially deposited in the Southern Swamp and its surroundings and melted there for several years), in fact, "dissolved" in the environment over the course of decades. Nevertheless, it is quite obvious that as the trajectory tilted below 24 km (up to a height of 7-8 km above the explosion epicenter), due to a sharp increase in braking, there was an intense release of cometary matter, resulting in the formation of a long cloud of evaporated hydrocarbons and air in the tail of each ice fragment. When the speed was reduced to 2.0 km/s, the wave front of the mixture burning in the tail caught up with the main volume of the cloud, which was stretched over 15-17 km (with a diameter of 4-5 km), followed by a so-called volumetric explosion that lasted about 5 seconds.  After 3 seconds, the shock wave reached the taiga, tearing off the crowns, breaking branches, and toppling trees.  After another 2-3 seconds, a mass of incandescent gases appeared, instantly scorching all the frontal surfaces of trees, shrubs, and moss that came into contact with it.

The first explosion was followed by three more strong and one less power  volumetric explosions,  increasing  the  destruction  of  the taiga.  Reflected  from  the  earth  surface, shock waves gave an additional impulse to the upward moving mass of the explosion products and the entrained mixture of unreacted hydrocarbon gases and water vapor. After 5-8 minutes, the huge mass of neutral products ejected by the explosion will be in the ionosphere, displacing the plasma to the periphery of the geomagnetic disturbance area.  <...>

Our average estimate of the total mass of the invaded TM was approximately 32 Mt (based on the average excess moisture content of ~ 0.7 cm of deposited water in the atmosphere over Mount Wilson during the post-disaster air mass 27 days of realization of this excess). In this case, nitrogen dioxide was used as a marker to determine the transition coefficient from the specific value we calculated (total content) to the initial (in the clouds of explosions) mass of the substance ($H_2O$). Model amount of NOx ($6 \cdot 10^{35}$ molecules) with our correction (a reduction of 12 times) was borrowed from the model presented in the work [Turco et al, 1982]. Our later estimates of the error of the model used in the aspect of interest to us have led us to the need to reduce the original model content of nitrogen oxides yet approximately 30 times (taking into account the possible additional production of NOx in clouds of incandescent gases after volumetric explosions). V. A. Bronshten [2000] later came to a similar conclusion. Since the calculated mass and the conversion factor are related by a linear equation, the calculated mass should be reduced by the same factor, resulting in approximately 30 Mt. If we take into account that the value of the explosion energy calculated using different methods differs by a factor of ~5 (from 9.5 to 50 Mt in TNT equivalent) among different authors, then the value of the mass we obtained and the accuracy of its determination can be considered acceptable. <...>

Post-catastrophic white nights at latitudes where they have not been observed before can be explained by the injection of approximately 0.15 Mt of fine-grained carbon and about 7.5 Mt of water into the mesosphere during explosions. This amount of explosive products is sufficient to form silvery clouds throughout the area from the explosion site to Bordeaux (France). In addition, the injection of a significant amount of explosive products and their filling of the ionosphere naturally caused a (five-hour) geomagnetic disturbance over the vast area of the disaster and the surrounding regions to the south and southeast.

In summary, it can be noted that the proposed model of the Tunguska phenomenon, before the Tunguska and post-catastrophic phenomena, satisfactorily explains the entire complex of known events associated with this phenomenon, and solves some problems that seemed previously insoluble."

These are the final words of the article. As follows from the publications, in order to explain the atmospheric optical anomalies before June 30, the authors proposed a version about a comet that was captured by the Earth's gravitational field and orbited the planet for several days.

Many researchers, and especially astronomers, are very skeptical about the authors' interpretation of the cometary infall as an explanation for the atmospheric optical anomalies.

The Author has devoted a lot of space to reviewing the above-mentioned articles on the interpretation of atmospheric transmission data, as they are crucial for understanding the causes of atmospheric optical anomalies associated with the Tunguska event.

## 3. Discussion

The first signs of the atmospheric optical anomalies had appeared already in the last decade of June [Vasilyev, 1998; Ol'khovatov, 2025a]. The anomalies reached their maximum on the night of June 30 to July 1. Beginning with July 1 they vanished gradually. The area of the atmospheric optical anomalies was limited by the Yenisei River in the east, line Tashkent-Stavropol-Sevastopol-Bordeaux in the south and the Atlantic shore in the west. The northern border merged with the area of "white nights" usual at that season. In 2008 it was discovered that at least in one place the bright nights also took place about fifty kilometers to the east of the Yenisei River, and by the way, on the nights preceded the Tunguska event [Ol'khovatov, 2020a].

Please, note that these anomalies are almost impossible to explain within the framework of the cometary interpretation, since:

1) The anomalies started in a weaker form a week (or maybe even earlier) before June 30, 1908.

2) In a few days, the tail (or coma) of the hypothetical "Tunguska" comet should have enveloped the entire Earth's surface. However, the anomalies were localized on the region of the Globe.

3) The cometary dust should have remained in the atmosphere for many weeks. However, the anomalies practically faded away in several days.

4) As can be seen from the above-mentioned articles on atmospheric transparency analysis, the Tunguska event did not insert much dust into the atmosphere.

5) The optical anomalies were also observed at relatively low latitudes, where they could not be caused by the scattering of sunlight by the cometary dust at night. For example, at the latitude of Tashkent, the sun only illuminates altitudes above ~700 km at dusk, where no dust from the hypothetical Tunguska Comet could have been suspended in the surrounding atmosphere.

6) The anomalies covered almost all layers of the atmosphere, from the troposphere to the ionosphere. The cometary interpretation has difficulty explaining how the cometary dust could penetrate all layers of the atmosphere at the same time and still be localized in a specific area.

7) The optical anomalies were not observed in the area of the Tunguska epicenter, where the greatest amount of comet dust should be expected to enter the Earth's atmosphere, but only appeared approximately 500 km west of the epicenter, at the longitude of the settlement of Sulomay (where they appeared before the Tunguska event, by the way).

The above arguments also apply to the asteroidal interpretation.

Here is what Alan Harris wrote [Harris, 1996]:

"Curiously, the anomalous "white nights" were reported as starting a full week before the impact, an observation that is hard to reconcile with even a comet since the Earth could not possibly spend a full week cruising through the tail. More likely, unusually intense auroral or noctilucent cloud activity unrelated to the impact occurred before the event."

In the Author's opinion the sky optical anomalies and the explosive event in Tunguska are related.

Now what the geophysical interpretation can say about the atmospheric optical anomalies of that period. From this point of view, the Tunguska event (as an event in the Central Siberia) and the atmospheric optical anomalies are different manifestations of the geophysical peculiarities of that period.

As for the increase in the glow of the night sky, the factors that probably have contributed to this increase have been discussed in [Ol'khovatov, 2025a]. Among the important factors were the atmospheric gravity waves. However the gravity waves can result also in rather spectacular noctilucent clouds, as, for example, it took place in August of 2025 - see ( https://spaceweather.com/archive.php?view=1&day=26&month=08&year=2025 ). In 1982 V.A. Romejko (or Romeiko) wrote [Romejko, 1982] that according to his opinion, the explosion of the Tunguska body (being a source of the internal gravity waves) generated additionally the noctilucent clouds over Eurasia, where conditions for their existence already were present. The Author would like to add that this point concerns also any other source of the gravity waves at those times.

The next important factor is the presence of a large amount of water vapor in the atmosphere. The author has already pointed out the increased volcanic activity before the Tunguska event [Ol'khovatov, 2003]. The second, and perhaps more important, aspect is the likely increase in endogenous hydrogen degassing [Ol'khovatov, 2025e] which could result in increased water vapour in the mesosphere.

The increased volcanic activity [Ol'khovatov, 2003] before the Tunguska event also added some dust into the atmosphere, which probably was interpreted in [Kondratyev et al., 1988] and the following works by the group of the authors as the "Pre-Tunguska meteoroid" [Nikol'skii et al., 2012].

An analysis of fluctuations in solar activity revealed some peculiarities in the solar cycle in which the Tunguska event occurred [Chirkov, 1986] (geomagnetic activity also had some peculiarities in the solar cycle and inside the year 1908 [Chirkov, 1986; Dmitriev, 1988]).

There was some increase of solar activity in the late June - early July, 1908

[Ol'khovatov, 2025a]. The increase of solar activity in late June and early July of 1908 may also have led to an increase in atmospheric aerosol concentrations. For example, according to [ Korshunov and Zubachev, 2021], solar proton events with a delay of 3–8 days are followed by an increase of 20% to 70% in aerosol lidar backscattering in the lower stratosphere. According to [ Korshunov and Zubachev, 2021] this effect has been shown to occur predominantly during the transport of stratospheric air to the observation point from the region of high latitudes. The latter may explain why the reported "long-twilight" optical anomalies of 1908 were often gravitated northwards. Aerosol was formed due to solar particles, and especially protons which penetrated deeper in the atmosphere. Solar particles ionize the air, forming ions, which then lead to the formation of aerosol in the atmosphere.

It is reasonable to add that besides the solar proton events, solar X-ray flares can produce ionization in the stratosphere [Murase et al., 2023], and so to form aerosol. However in the lower stratosphere, their effectiveness decreases dramatically.

The atmospheric optical anomalies of the time of interest were also manifested as changes in positions of the neutral points of Arago and Babinet [Vasil'ev et al., 1965; Vasilyev, 1998].

Volcanic aerosol in the atmosphere results in the shift of the Arago and Babinet neutral points of sky polarization away from their normal positions on the sky [Stothers, 1996], which is partly in agreement with the case of the Tunguska event (but there was some difference also). In the case of the ice crystals in the atmosphere, the situation seems similar to volcanic aerosol, at least, for the neutral point of Arago - the ice crystals have shifted the Arago point (as mentioned in the article [Fitch and Coulson, 1983]) by several degrees away from the antisolar point, and the effects of ice crystal scattering mimic the effects of aerosol scattering [Fitch and Coulson, 1983], however the observed ice crystals in [Fitch and Coulson, 1983] seem to be larger than in the noctilucent clouds.

An idea that there is a connection between solar activity and positions of the neutral points of Arago and Babinet was put forward already in 1893 [Maurer, 1915]. Observations in [König, 1961] are in agreement with this idea. The Author would like to add some comments to the conclusion of the author of [König, 1961] due to the data published in 1990. In [König, 1961] there was a shift of the Arago and Babinet neutral points of sky polarization on Aug. 30, 1957 away from their positions compared with the positions on Sept. 6, 1957 (i.e. the solar distances were increased on Aug.30). According to [Shea and Smart, 1990] there were 2 solar proton events on Aug. 29, and the second one was peaking on Aug.30. Also according to [König, 1961] an aurora borealis was visible on the night of Aug. 29/30, 1957, albeit only slightly. The proton event preceding to the measurements on Sept. 6, 1957 was only on Sept. 2 [Shea and Smart, 1990]. However on Sept.6, 1957 there were unfavorable weather conditions for observations. It can be seen from Fig.1 of [König, 1961]  that the Arago positions on Sept. 1-6, 1958 were not far from those of Aug.30, 1957 when the Sun was above the horizon (the data for Aug.30, 1957 is partly absent when the Sun is below the horizon).

According to [König, 1961] the solar distances of the Babinet points on August 30, 1957 approach the high value of those on September 1-6, 1958, and are on average only 1.2° lower.  Also according to [König, 1961], the solar distances of the Babinet points on September 5, 1958 in the morning, immediately after the appearance of the aurora borealis are on average 1.4° greater than on the previous evening. According to [Shea and Smart, 1990] the solar proton event preceding observation of Sept. 1-6, 1958 was on Aug. 26, 1958. It was written in [König, 1961] that from September 3-5, 1958, an extraordinarily strong aurora borealis occurred, which was visible throughout Central Europe from the North Sea coast to Munich and Nuremberg, in Normandy and in America.

The article [König, 1961] concludes that thus there is possibility ("die Möglichkeit" in German), that the large solar distances on September 1-6, 1958 and on August 30, 1957 were caused by the corpuscular radiation, which also caused the aurora at this time.

The author of [König, 1961] wrote about "possibility", as just 2 cases are not enough for a solid statement. Here the Author adds one more case in favor of the statement. In [Longtin and Volz, 1986] there is data on positions of the neutral points of Arago and Babinet in September and October of 1969. According to [Shea and Smart, 1990] there were 2 solar proton events on Sept. 25, and Sept. 27 (1969) and the second one was peaking on Sept.28. There is data in [Longtin and Volz, 1986] on positions of the neutral points of Arago and Babinet for those times.  It can be seen from the data that there was some upsurge in the distances for the neutral points of Arago and Babinet comparing with the several preceding days at least. Unfortunately clouds appeared since Sept.30, so duration of the upsurge is hard to estimate.

In the opinion of the Author, the König's word "possibly" can be changed to "probably". But of course, further research is needed for a reliable conclusion.

The observational data is not very clear regarding relations between noctilucent clouds and solar proton events. For example, in [von Savigny et al., 2007] immediately after the onset of the enhanced solar particle precipitation on January 16, 2005, a severe decrease was observed in the noctilucent clouds occurrence rate. However in [Bardeen et al., 2016] it was shown that the very strong solar proton event of 23 – 30 January 2012 produced just a small effect. Discussions of the physical mechanisms is beyond the scope of this paper. So if there was a solar proton event about June 30, 1908 indeed, as it was admitted by the Author in [Ol'khovatov, 1997], its negative effect on the noctilucent clouds could be just a small one.

Anyway, there is an alternative (or an addition?) to the alleged role of the solar activity. In [Ol'khovatov, 2025f] the Author presented arguments that the Tunguska event was associated with electrical discharges (and a thunderstorm accompanied the Tunguska event). The electrical discharges could result in formation of aerosol in the atmosphere. Indeed, in [Wang et al., 2021] it was discovered that numerous ultrafine particles were formed during the lightning events, leading to increases in nucleation and Aitken mode aerosols by 18.9 and 5.6 times, respectively.

The electrical discharges can also result in production of nitrogen oxides and acid rains, as shown in [Railsback, 1997].

One more aspect is that there are electric discharges emerging from the cloud top to the bottom of the ionosphere, like, for example, the "gigantic jets" [Rycroft and Harrison, 2012]. According to eyewitness accounts, the size of the glowing area was quite large during the Tunguska event. Here's what A.K. Kokorin (an observer of a meteorological station in Kezhma, which was about 215 km to SSW from the epicenter) wrote in the log-book of the meteo-station, translated by A.O.:

"June 30…new calendar at 7 o'clock in the morning, two huge fiery circles appeared in the north; after 4 minutes from the beginning of the appearance, the circles disappeared; soon after the disappearance of the fiery circles, a strong noise was heard, similar to the noise of the wind, which went from north to south; the noise lasted about 5 minutes. Then followed the sounds and crackling, similar to the shots from huge guns, from which the frames trembled. These shots lasted for 2 minutes, and after them there was a crackling sound, similar to a shot from a gun. These latter lasted 2 min. Everything that happened was under a clear sky".

The Author wrote a little bit about him in [Ol'khovatov, 2023b]. According to his description, the upper boundary of the glowing circles was at least several tens of kilometers above the ground. Under such conditions, it is possible that the ionosphere was also affected, as evidenced by the geomagnetic disturbance associated with the Tunguska event.

So the topic is of possible interest of researchers working on the problem of lithosphere - atmosphere - ionosphere coupling. The Author incline to think that probably the deep layers of the Earth were involved too. Anyway the influence of the Tunguska event on the global atmospheric electric circuit (about the circuit - see [Rycroft and Harrison, 2012], for example) remains to be explored.

A separate aspect is finding similar or even more energetic events in the history of the planet. It would be interesting, for example, to consider from this perspective some of the processes associated with the onset of the Younger Dryas and discussed, for example, in the article [Kennett et al., 2025]. The Author hopes that some researchers will be interested to do this.

## 4. Conclusion

The geophysical interpretation of the Tunguska event is in agreement with the atmospheric optical anomalies at the qualitative level at least. Quantitative explanation is currently difficult due to limited observational data and the lack of clarity in the details of physical phenomena. Neither the asteroidal nor the cometary interpretation of the Tunguska event is consistent with the anomalies.

The general conclusion is that the Tunguska event was a very complex phenomenon. Research of the Tunguska event requires the participation of experts in various fields.

## ACKNOWLEDGEMENTS

The Author wants to thank the many people who helped him to work on this paper, and special gratitude to his mother  - Ol'khovatova Olga Leonidovna (unfortunately she didn't live long enough to see this paper published...), without her moral and other diverse support this paper would hardly have been written.